# Local view of superconducting fluctuations

Shai Wissberg, Aviad Frydman, and Beena Kalisky
*Department of Physics and Institute of Nanotechnology and Advanced Materials, Bar-Ilan University, Ramat-Gan 5290002, Israel*

Superconducting transitions are driven by thermal fluctuations close to the transition temperature, $T_c$. These fluctuations are averaged out in global measurements, leaving imprints on susceptibility and resistance measurements. Here, we use a scanning superconducting quantum interference device to image thermal superconducting fluctuations in Nb, a conventional BCS superconductor. We observe fluctuations in both space and time which manifest themselves as grains of weaker and stronger diamagnetic response, exhibiting telegraph-like noise as a function of time. Local fluctuations are also found in the imaginary component of the susceptibility demonstrating that the local vortex dissipation can also be used as a probe of the fluctuations. An important outcome of our measurements is the observation and realization that the susceptibility decrease to zero as the temperature is raised towards Tc always occurs in quantized steps irrespective of the sample geometry. The technique described here is expected to be a useful tool for studying the nature of fluctuations in more complex superconductors, providing important information on critical properties such as fluctuation length, time scales, and local viscosity. © 2018 Author(s). All article content, except where otherwise noted, is licensed under a Creative Commons Attribution (CC BY) license (http://creativecommons.org/licenses/by/4.0/). https://doi.org/10.1063/1.5037702

Superconductors are characterized by a sharp transition at a critical temperature, Tc, where electrons pair, the resistivity drops to zero, and diamagnetism sets in. However, the presence of fluctuations broadens this transition. These fluctuations are well described within the framework of the Ginzburg-Landau theory[1] yielding length and time scales which diverge at $T_c$

$$\xi \propto \sqrt{\frac{T_c}{T_c - T}}; \quad \tau = \frac{\hbar\pi}{8K_B(T - T_c)}. \qquad (1)$$

Although the presence of superconducting fluctuations is well established and their thermodynamic properties are well understood,[2,3] they have never been directly imaged. Here, we present an experiment designed to study the local aspects of superconducting fluctuations. Our measurements enable us to observe fluctuations of the diamagnetic response in both time and space in a very narrow temperature window around the critical temperature. We demonstrate the ability to investigate superconducting fluctuations on a conventional BCS superconductor, and suggest it as a powerful tool for studying the local nature of criticality in more complex superconductors.

For this purpose, we use a scanning superconducting quantum interference device (SQUID) with excellent magnetic sensitivity[4,5] to probe the local diamagnetic response from a 150 nm thick Nb film. We record the magnetic response of the sample (units of $\Phi_0$) to an alternating magnetic field applied by an on-chip field coil, concentric with the SQUID's sensing area, the pickup loop [see Fig. 1(a)]. This measurement detects the diamagnetic response of a superconductor to a magnetic excitation, in units of $\Phi_0/A$ (flux reading in the pickup loop divided by the current in the field coil). Unlike magnetization measurements, we probe the local response of a film to an applied magnetic excitation at a chosen frequency. The a.c. measurements allow extraction of the local full response function[6] $\chi = \chi' + i\chi''$, where the in-phase component, $\chi'$, is generally associated with the superfluid density, $\rho_S$, and the out-of-phase component, $\chi''$, measures the viscosity of the flux-flow, related to dissipation.[7] While the former drops from the full diamagnetic response at low temperature to zero as the temperature approaches $T_c$, the latter is measurable only at a narrow temperature region near $T_c$.[8–10]

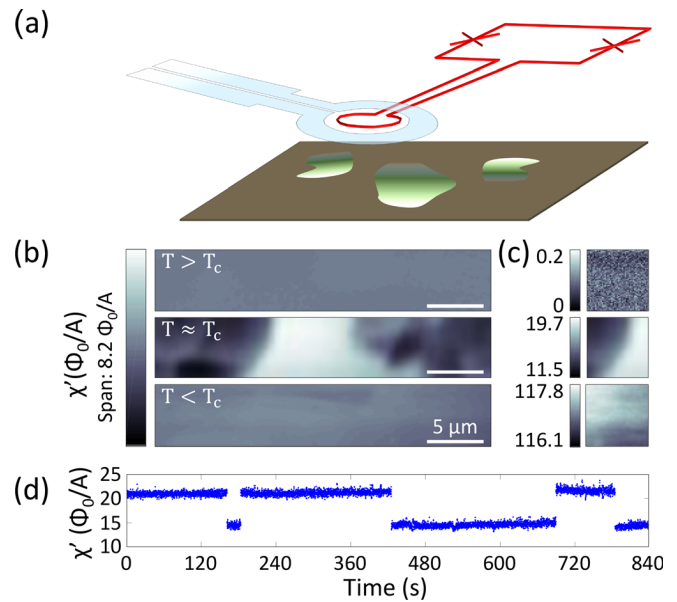

FIG. 1. Scanning SQUID measurements of superconducting fluctuations. (a) A sketch, not to scale, showing how the scanning SQUID measures the local diamagnetic response. The SQUID loop (red), with the inner diameter of 1.8 μm, which sets our separation limit, is surrounded by a single turn loop, the field coil (light blue), used to apply local magnetic field. The sample's response to the local magnetic field is recorded using the pickup loop and presented in units of flux over the current in the field coil, $\Phi_0/A$. (b) and (c) Scanning SQUID susceptibility scans of a certain area at different temperatures, plotted with the same span (b) and with individual color spans (c). We define $T_c$ as the temperature above which no local diamagnetic response is detected. Here, $T_c = 8.986$ K. Above $T_c$, at $T = 8.99$ K, no diamagnetic response is detected and the image taken represents the SQUID's noise. Near $T_c$, at $T = 8.925$ K, different levels of diamagnetic response are measured, showing strong modulations relative to the ones measured below $T_c$, at $T = 8.734$ K. (d) Susceptibility at a specific point recorded as a function of time. The susceptibility signal fluctuates between two different values. The measurement temperature was 8.921 K.

Employing the scanning SQUID, we image variations in the superconductivity close to $T_c$ in two ways. The first is by rastering the excitation coil over the sample, and recording the measured flux close to the sample, as a function of position, thus generating maps of the local diamagnetic response on a scale of microns.[11,12] Figure 1 shows that above $T_c$ no diamagnetism is detected, and the recorded image reflects the noise of our measurement. Below $T_c$ the diamagnetic response is again uniform, interrupted only by sample inhomogeneities. In a narrow region near $T_c$, the susceptibility maps reveal a rich structure, arising from superconducting fluctuations [Figs. 1(b) and 1(c)]. The second way we probe fluctuations is by parking the SQUID at a certain location and measuring $\chi$ as a function of time. This way, we continuously check, at a chosen frequency (923 Hz), what is the local ability to shield the applied magnetic field. This method is different than a continuous d.c. measurement of magnetism which detects flux flow of mobile vortices at elevated temperatures.[13] Close to $T_c$ the time-traces show a telegraph-like noise, such as that of Fig. 1(d). The discrete values in the telegraph-like time trace describe a discrete change in ability of the superconductor to eliminate the applied field.

Fixing the position of the SQUID enables the measurement of local diamagnetic response, $\chi$, as a function temperature [Fig. 2(a)]. The temperature of the SQUID is decoupled from the temperature of the sample, allowing high sensitivity SQUID measurements of the sample below and above its $T_c$. As expected,[14] $\chi'$ drops to zero at $T_c$. Interestingly, close to $T_c$ the decrease in $\chi'$ with increasing T occurs in jumps between discrete levels as depicted in the inset of Figs. 2(a) and 2(c). In Fig. 2(b), we use images of a single vortex and its dependence on the penetration depth, $\lambda$, to show that the amplitude of the steps is consistent with the signal from a vortex with a large $\lambda$. This result, of a step-like decrease in the susceptibility signal, is a natural manifestation of the way supercurrents shield magnetic field. When the magnetic field is applied to a superconductor, the supercurrents flow in closed paths, maintaining flux quantization.[15,16] Therefore, sensitive susceptibility measurements should detect discrete levels originating from the addition or subtraction of an individual fluxoid. Our measurements demonstrate that these effects are not limited to obvious ring geometries where the flux is naturally quantized[13,17] but are a general feature of susceptibility measurements on any superconductor, since the flowing screening currents involve $2\pi$ phase quantization. Figure 2(c) shows that the amplitude of these jumps decreases with increasing the temperature, in agreement with the conjecture that the size of the jumps is due to a change of one fluxoid.

The steps in the susceptibility signal appear only in a narrow region of few mK. In this temperature window, we observe susceptibility jumps not only as a function of temperature but also as a function of time, as seen in Fig. 3. These result from thermal fluctuations in the very near vicinity of the transition.[20] On the average, the typical fluctuation time, $\tau$, is expected to diverge at the transition according to Eq. (1). Locally, however, the situation is more complicated. The nucleation and growth of regions of weaker and stronger superconducting islands leads to a wide spread of characteristic times even at a fixed temperature.

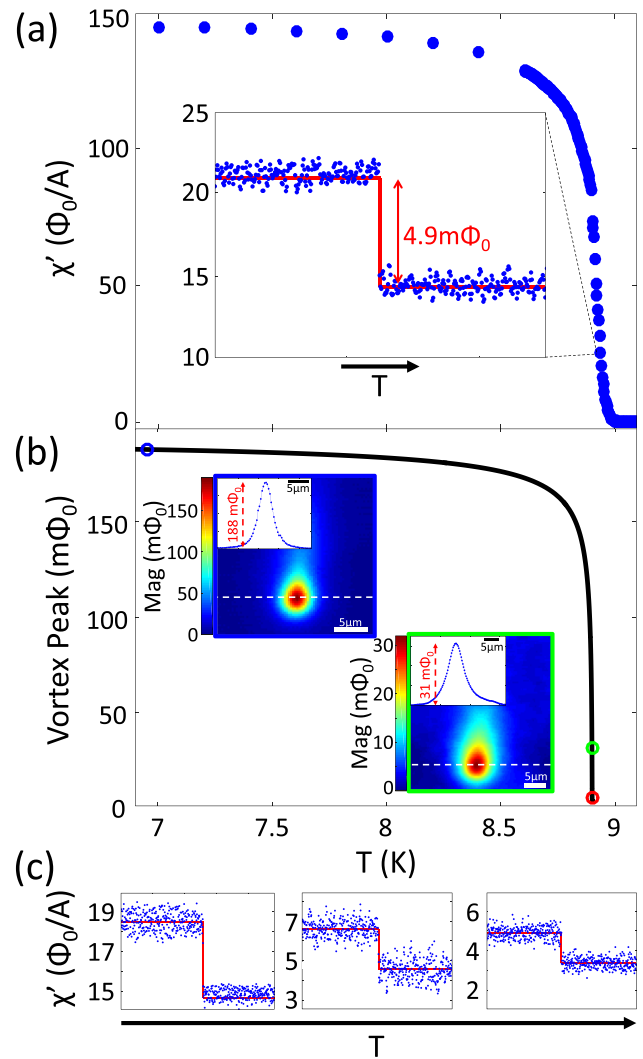

FIG. 2. Measurements of the local susceptibility. (a) The local susceptibility measured far from the edges of the sample as a function of temperature. Inset: Close to $T_c$, the susceptibility decreases in discrete steps. We convert to flux ($m\Phi_0$), by multiplying $\chi'$ ($\Phi_0/A$) by the current driven through the field coil. The system temperature was set to 8.921 K and was allowed to naturally drift up in a regime of a few mK. (b) The peak of a vortex signal as a function of temperature,[18,19] as imaged by the SQUID. The blue and green circles are values measured by imaging individual vortices at the relevant temperatures. Two insets show the flux image recorded by scanning SQUID over an individual vortex. Each vortex contains one $\Phi_0$, extends over a distance determined by $\lambda$, and is resolution limited by the size of the probe and the height of the scan, which do not change as a function of temperature. Therefore, the amplitude of the flux signal decreases with increasing $\lambda$, reaching $m\Phi_0$ for $\lambda$ of 6.5 $\mu$m. The amplitude of the steps we observe is consistent with the amplitude of a phase change of $2\pi$, a fluxoid. Profiles extracted from the locations marked by white lines are also presented in the insets. The red circle corresponds to the susceptibility signal depicted in the inset of (a). (c) Several discrete steps in the susceptibility taken by setting the system temperature to 8.919 K, 8.932 K, and 8.934 K, respectively, and allowing it to naturally drift up in a regime of a few mK. The measured susceptibility becomes weaker as the temperature is increased, as well as the amplitude of the steps, with measured jump amplitudes of 3.8$\Phi_0/A$, 2$\Phi_0/A$, and 1.5$\Phi_0/A$.

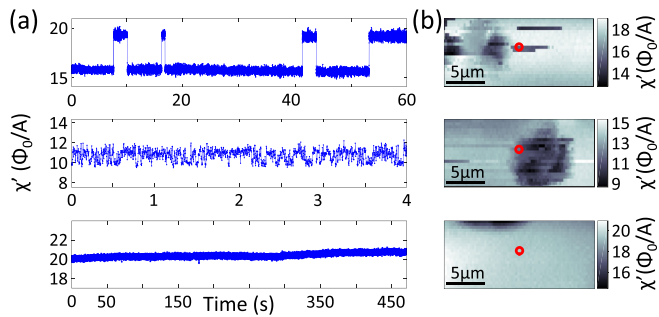

FIG. 3. Thermal fluctuations as recorded in the susceptibility signal. (a) Telegraph-like noise graphs of $\chi'$ taken at three different locations, positions on the sample shown in (b). The measurement temperature was 8.927 K (top), 8.934 K (middle), and 8.924 K (bottom). The time between $\chi'$ switches is very diverse, ranging from a few tens of milliseconds (middle) to a few minutes. Areas that are far from the superconducting transition show no switching events (bottom). (b) $\chi'$ maps, with red dots marking the location at which the time traces, were measured. The horizontal streaks in the image are due to fluctuations happening mid-scan.

Note that the temperature uncertainty of our measurement is 1 mK. Depending on the location, we observe characteristic switching times that range from tens of milliseconds to tens of seconds for a single sample as depicted in Fig. 3. In certain areas, we do not observe switching events (Fig. 3), presumably because our applied field is not sufficient to induce a vortex or that the switching time is longer than the duration of our measurements. All in all, the range of $\tau$ in this sample is probably much larger than tens of milliseconds, but we are limited by our excitation frequency on one hand and by the measurement duration on the other.

The wide spread of $\tau$ leads to the presence of fluctuations in space which can be detected by our probe. Rastering the SQUID upon the sample reveals a rich spatial distribution of $\chi'$, such as that seen in Fig. 4(a), which shows susceptibility maps ($20 \times 8\ \mu m^2$) close to the transition, featuring regions of weaker and stronger response that develop with temperature. Such spatial variation of susceptibility values near $T_c$ is often related to local $T_c$ variations in unconventional superconductors.[21] Choosing to study a basic BCS superconductor minimizes the influence of the inevitable spatial distribution of $T_c$ and allows observation of dynamic fluctuations.

Figure 4(a) also shows $\chi''$ maps at different temperatures. These images, which reflect the local distribution of the dissipative component of the susceptibility related to the vortex viscosity, exhibit the structure that is similar to that of $\chi'$. Similarly, Fig. 4(b) shows that the time traces of $\chi''$ exhibit telegraph like noise that resembles those of $\chi'$. This demonstrates that the imaginary component of $\chi$ can also be used as a tool to analyze the local aspects of superconductivity near criticality.

To conclude, we demonstrate that the scanning SQUID provides a unique means to study local fluctuation phenomena close to criticality which is not accessible in a global measurement. Apparently, even for the conventional BCS superconductor, Nb, the measurements provide a number of unforeseen (though logical) results such as the quantized decrease in susceptibility to zero close to the transition with step magnitudes of single fluxoids, and the spread of fluctuation times observed in a single sample. We expect this method to serve as a powerful analysis tool for fluctuations in more complicated systems. One example is the study of quantum fluctuations close to a quantum critical point.[22]

We thank A. Sharoni for providing the superconducting samples, N. Vardi for assistance with the measurements, E. Persky for assistance with simulations, and Y. Yeshurun and A. Shaulov for helpful discussions. S.W., A.F., and B.K. acknowledge COST Action CA16218. S.W. and B.K. were supported by the European Research Council Grant No. ERC-2014-STG-639792. A.F. was supported by Israel U.S. binational foundation Grant No. 2014325.

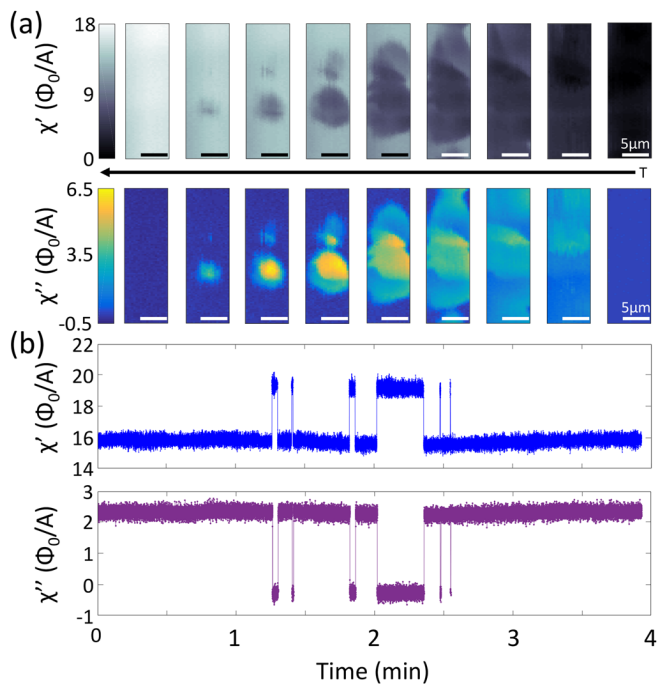

FIG. 4. Relations between $\chi'$ and $\chi''$, as a function of temperature and time. (a) Maps of the real (upper) and imaginary (lower) terms of the susceptibility, taken by setting the system temperature to 8.934 K and letting it naturally drift down in the regime of a few mK. Black arrow represents the cooldown direction. (b) $\chi'$ (blue) and $\chi''$ (purple) measurements taken simultaneously on a specific point as a function of time, at 8.919 K. Both signals fluctuate between two distinct values, changing simultaneously. All switching events we observed occurred simultaneously in the in-phase and out-of-phase components.

[1]V. L. Ginzburg and L. D. Landau, *On Superconductivity and Superfluidity* (Springer-Verlag Berlin Heidelberg, 2009), pp. 113–137.
[2]R. E. Glover, Physica **55**, 3 (1971).
[3]A. Larkin and A. Varlamov, *Theory of Fluctuations in Superconductors* (OUP, Oxford, 2009).
[4]J. R. Kirtley, M. B. Ketchen, K. G. Stawiasz, J. Z. Sun, W. J. Gallagher, S. H. Blanton, and S. J. Wind, Appl. Phys. Lett. **66**, 1138 (1995).
[5]J. R. Kirtley and J. P. Wikswo, Annu. Rev. Mater. Sci. **29**, 117 (1999).
[6]R. B. Goldfarb, M. Lelental, and C. A. Thompson, *Magnetic Susceptibility of Superconductors and Other Spin Systems* (Springer US, 1991), p. 4980.
[7]J. V. de Vondel, J. Van de Vondel, B. Raes, and A. V. Silhanek, *Superconductors at the Nanoscale* (Berlin, Boston: De Gruyter, 2017).
[8]P. Das, C. V. Tomy, S. S. Banerjee, H. Takeya, S. Ramakrishnan, and A. K. Grover, Phys. Rev. B Condens. Matter **78**, 214504 (2008).
[9]S. Casalbuoni, E. A. Knabbe, J. Ktzler, L. Lilje, L. von Sawilski, P. Schmser, and B. Steffen, Nucl. Instrum. Methods Phys. Res. A **538**, 45 (2005).

[10]R. B. Goldfarb, A. F. Clark, A. I. Braginski, and A. J. Panson, Cryogenics **27**, 475 (1987).

[11]M. E. Huber, N. C. Koshnick, H. Bluhm, L. J. Archuleta, T. Azua, P. G. Bjrnsson, B. W. Gardner, S. T. Halloran, E. A. Lucero, and K. A. Moler, Rev. Sci. Instrum. **79**, 053704 (2008).

[12]B. W. Gardner, J. C. Wynn, P. G. Bjrnsson, E. W. J. Straver, K. A. Moler, J. R. Kirtley, and M. B. Ketchen, Rev. Sci. Instrum. **72**, 2361 (2001).

[13]J. R. Kirtley, C. C. Tsuei, V. G. Kogan, J. R. Clem, H. Raffy, and Z. Z. Li, Phys. Rev. B: Condens. Matter Mater. Phys. **68**, 214505 (2003).

[14]F. Gomory, Supercond. Sci. Technol. **10**, 523 (1997).

[15]B. S. Deaver and W. M. Fairbank, Phys. Rev. Lett. **7**, 43 (1961).

[16]R. Doll and M. Nbauer, Phys. Rev. Lett. **7**, 51 (1961).

[17]J. E. Lukens and J. M. Goodkind, Phys. Rev. Lett. **20**, 1363 (1968).

[18]J. R. Kirtley, C. C. Tsuei, K. A. Moler, V. G. Kogan, J. R. Clem, and A. J. Turberfield, Appl. Phys. Lett. **74**, 4011 (1999).

[19]V. G. Kogan, Phys. Rev. B: Condens. Matter Mater. Phys. **68**, 104511 (2003).

[20]W. J. Skocpol and M. Tinkham, Rep. Prog. Phys. **38**, 1049 (1975).

[21]C. W. Hicks, T. M. Lippman, M. E. Huber, J. G. Analytis, J.-H. Chu, A. S. Erickson, I. R. Fisher, and K. A. Moler, Phys. Rev. Lett. **103**, 127003 (2009).

[22]A. Kremen, H. Khan, Y. L. Loh, T. I. Baturina, N. Trivedi, A. Frydman, and B. Kalisky Imaging quantum fluctuations near criticality (unpublished).